\documentclass[structabstract]{aa}  
\usepackage{natbib}
\bibpunct{(}{)}{;}{a}{}{,} 
\usepackage{graphicx}
\usepackage{txfonts}
\usepackage{lscape}
\newcommand{\rosat}{{\sl ROSAT}}
\newcommand{\chandra}{{\sl Chandra}}
\newcommand{\xmm}{{\sl XMM-Newton}}
\newcommand{\n}[1]{$N_\mathrm{#1}$}

\def\hii{H\,{\sc ii}}

\begin{document}
   \title{\xmm\ observations of the superbubble in N\,158 in the LMC}

   \subtitle{}

   \titlerunning{\xmm\ observations of N\,158}

   \author{Manami Sasaki
          \inst{1}
          \and
          Dieter Breitschwerdt\inst{2} 
          \and
          Verena Baumgartner\inst{3} 
          \and
          Frank Haberl\inst{4} 
          }

   \institute{Institut f\"ur Astronomie und Astrophysik, 
              Universit\"at T\"ubingen,
              Sand 1, 
              D-72076 T\"ubingen, Germany,
              \email{sasaki@astro.uni-tuebingen.de}
         \and
              Department of Astronomy and Astrophysics,
              Berlin Institute of Technology,
              Hardenbergstr.\ 36,
              D-10623 Berlin, Germany
         \and
              Institut f\"ur Astronomie,
              Universit\"at Wien,
              T\"urkenschanzstr.\ 17,
              A-1180 Vienna, Austria
         \and
              Max-Planck-Institut f\"ur extraterrestrische Physik, 
              Giessenbachstra{\ss}e
              D-85748 Garching, Germany
             }

   \date{Received October 05, 2010; accepted February 01, 2011}

 
  \abstract
   {}
   {We study the diffuse X-ray emission observed in the field of view of the pulsar
B\,0540--69 in the Large Magellanic Cloud (LMC) by \xmm. We want to understand the
nature of this soft diffuse emission, which coincides with the superbubble in the 
\hii\ region N\,158, and improve our understanding of the evolution of superbubbles. 
}
   {We analyse the \xmm\ spectra of the diffuse emission. Using the parameters obtained
from the spectral fit, we perform calculations of the evolution of the superbubble.
The mass loss and energy input rates are based on the initial mass function (IMF)
of the observed OB association inside the superbubble.}
   {The analysis of the spectra shows that the soft X-ray emission arises from hot 
shocked gas surrounded by a thin shell of cooler, ionised gas. We show that the stellar 
winds alone cannot account for the energy inside the superbubble, but the energy release 
of 2 -- 3 supernova explosions in the past $\sim$1~Myr provides a possible explanation.
} 
   {The combination of high sensitivity X-ray data, allowing spectral analysis, 
and analytical models for superbubbles bears the potential to reveal the 
evolutionary state of interstellar bubbles, if the stellar content is known.}


   \maketitle
%

\section{Introduction}

Early observations in the radio and the optical have shown that the
interstellar medium (ISM) in the Milky Way mainly consists of cool clouds 
($T \la 10^{2}$~K) of neutral hydrogen embedded in warm ($T \simeq 10^{4}$~K)
intercloud medium of partially ionised hydrogen. Since the 1970s, observations 
in the ultraviolet (UV) and X-rays showed the presence of hot gas at coronal   
temperatures ($T \simeq 10^{5-6}$~K) in the ISM.      
The heat source of the ISM are massive OB stars, which inject  
energy through their radiation, stellar winds, and finally by supernova     
explosions.                                                                 
As these processes are often correlated in space and time, superbubbles with 
sizes of typically $100 - 1000$~pc are created in the ISM.   
Therefore, supernova remnants (SNRs) and superbubbles are among the prime 
sources that control the
morphology and the evolution of the ISM, and their observation is
of key interest to understand the galactic matter cycle.
However, they radiate copiously in the soft X-rays below 2~keV, an energy range 
that is difficult to study in the Milky Way because of absorption by the 
Galactic disk. 

The LMC, which is a dwarf irregular, but shows
indications for spiral structures, is one of the closest neighbours of our
Galaxy. Its proximity with a distance of 48~kpc \citep{2006ApJ...652.1133M} 
and modest extinction in the line of sight (average Galactic foreground
$N_{\rm H} = 1.6 \times 10^{21}$~cm$^{-2}$) make it the ideal laboratory for
exploring the global structure of the ISM in a galaxy.
The well-known and best studied extended emission region in                
the LMC is the 30 Doradus region and the region south of it, which harbor star 
formation sites, superbubbles, and SNRs. \rosat\ data of the superbubbles
in the LMC have been studied in detail by, e.g., 
\citet{1995ApJ...450..157C} and \citet{2001ApJS..136..119D}.

N\,158 \citep{1956ApJS....2..315H} 
is one of the \hii\ regions in these active regions of the LMC. It is 
elongated in the north-south direction and consists of a superbubble in the 
north and a more compact bright region in the south. It is known to host two 
OB associations LH\,101 and LH\,104 \citep{1970AJ.....75..171L}. While 
LH\,101 in the southern part of N\,158 seems to power the very bright region in 
H$\alpha$, LH\,104 is found in the superbubble in the northern part of N\,158 
and is dominated by B stars \citep{1992A&AS...92..729S}, mainly consisting of 
a young population with an age of 2 -- 6~Myr \citep{1998A&AS..130..527T}.
\citet{2001ApJS..136..119D} have analyzed the \rosat\ data
and suggested that the X-ray emission seen at the position of N\,158 
is associated with the \hii\ region.
N\,158 is located near the X-ray bright pulsar (PSR) in the LMC 
B\,0540$-$69, which has been observed for calibration purposes for the 
X-ray Multi-Mirror Mission \xmm\ 
\citep{2001A&A...365L...1J,2000SPIE.4012..731A}. The field of view of the 
European Photon Imaging Cameras 
\citep[EPICs,][]{2001A&A...365L..18S,2001A&A...365L..27T} of these observations
when performed in Full Frame mode, covers the northern part of 
N\,158 and allows to study the X-ray emission from the superbubble. 

\section{Data}

The pulsar B\,0540$-$69 in the LMC is a Crab-like pulsar with a pulsar wind 
nebula (PWN), which has been spatially resolved and studied with the \chandra\
X-ray Observatory \citep{2007ApJ...662..988P}. 
To study the diffuse emission in the vicinity of B\,0540$-$69, we have
chosen those observations for which the EPICs were operated in Full Frame mode.
The observation IDs are 0117510201, 0117730501, and 0125120101. 
The observations were all carried out using the medium filter. 
Starting from the observational data files (ODFs), the data are processed with
the \xmm\ Science Analysis System (XMMSAS) version 10.0.0.
For EPIC PN, only single and double pattern events are used,
whereas for the MOS1 and 2, singles to quadruples are selected.
The exposure times that we obtain after 
filtering out the time intervals with background flares are listed in 
Table \ref{obstab}.

\begin{table}
\begin{center}
\caption{\label{obstab} \xmm\ data used for the analysis.
All the analysed data were obtained in Full Frame mode using the medium filter.}
\begin{tabular}{clcc}
\hline\hline
Obs.\ ID & EPIC & Start Date & Effective Exposure \\
 & & & [ksec] \\
\noalign{\smallskip}\hline\noalign{\smallskip}                       
01175102 & PN & 2000-02-11 & 8.3 \\
01175102 & MOS1,2 & & 3.5 \\
01177305 & PN & 2000-02-17 & 8.3 \\
01177305 & MOS1,2 & & 9.8 \\
01251201 & PN & 2000-05-26 & 29. \\
01251201 & MOS1,2 & & 27. \\
\noalign{\smallskip}\hline
\end{tabular}
\end{center}
\end{table}

\subsection{EPIC image}

\begin{figure*}
\centering
\includegraphics[width=0.8\textwidth,angle=0,clip=]{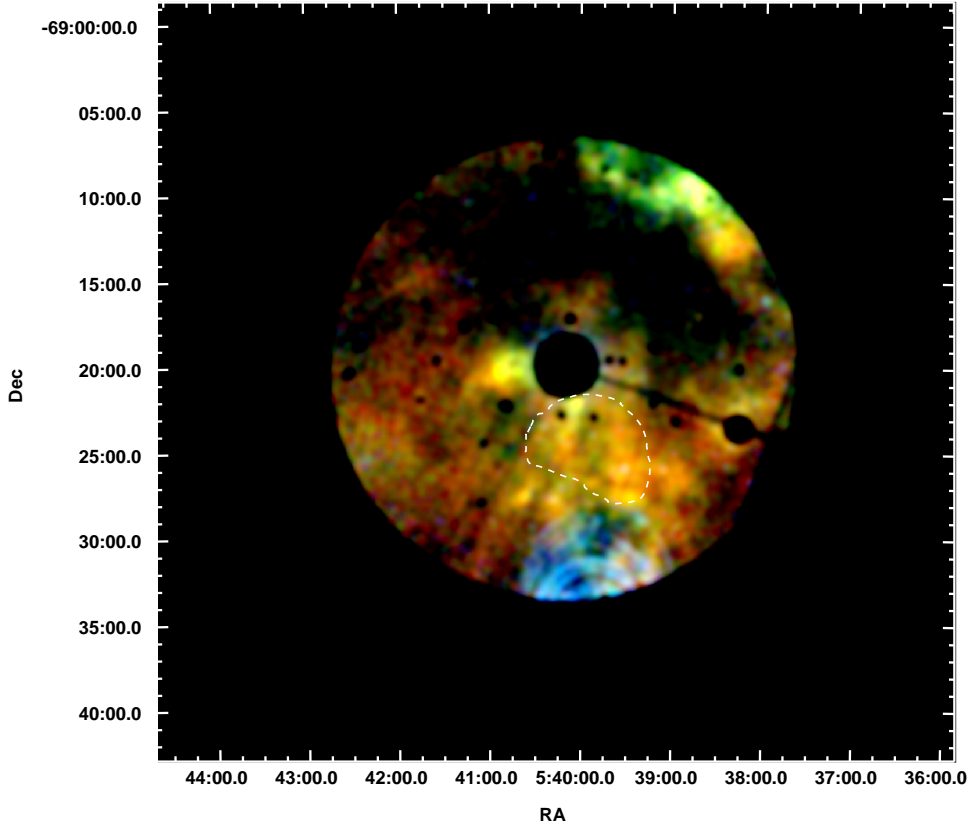}
\caption{\xmm\ EPIC mosaic image of the
  PSR\,B\,0540$-$69 and its surroundings 
  in true color presentation (red: 0.3 -- 0.8~keV,
  green: 0.8 -- 1.5~keV, blue: 1.5 -- 2.3~keV). The bright X-ray emission from
  PSR\,B\,0540$-$69 at $\sim$RA = 05$^{\sl h}$ 40$^{\sl m}$, 
  Dec~=~--69\degr~20\arcmin, other point sources, and the out-of-time events have 
  been removed from the data. The arc-shaped features
  in the south are caused by stray light from the bright X-ray source LMC X-1.
  The position of the superbubble in the \hii\ region N\,158 is shown with a 
  dashed line.}
\label{mosaic} 
\end{figure*}

\begin{figure*}
\centering
\includegraphics[width=0.49\textwidth,angle=0,clip=]{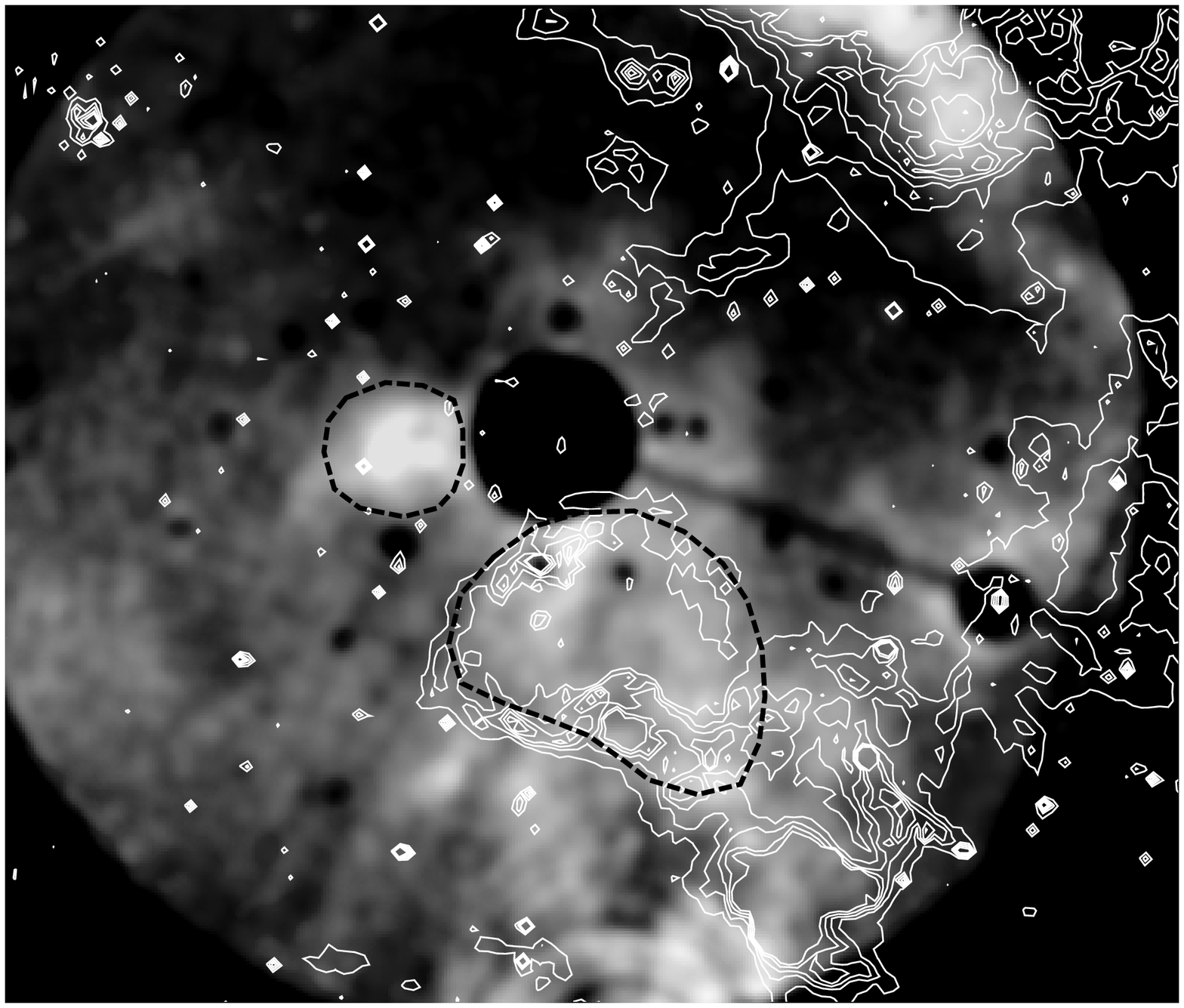}
\includegraphics[width=0.49\textwidth,angle=0,clip=]{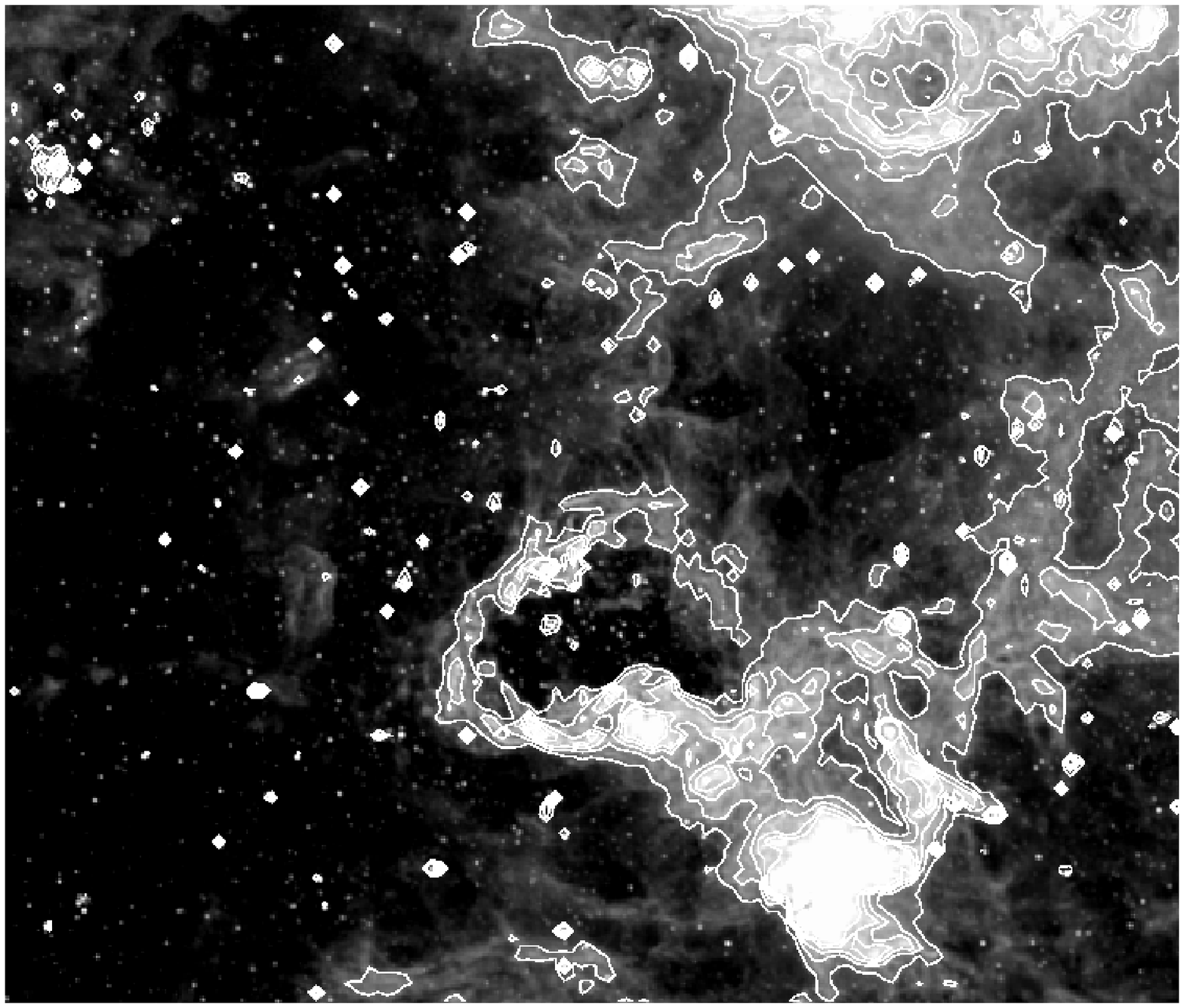}\\
\caption{A zoom in on the \xmm\ EPIC mosaic image with regions 
used for the spectral analysis (black dashed line) and H$\alpha$ contours 
(MCELS, left) and the H$\alpha$ image with the same contours (right). 
}
\label{hahi} 
\end{figure*}

After filtering out the background flares, we created a mosaic image out 
of the Full Frame mode data of EPIC PN, MOS1, and MOS2 for all three 
observations (Fig.\,\ref{mosaic}). To enhance the not so bright diffuse 
emission, we filtered out all point sources that have been found in a source 
detection routine as well as the so called out-of-time events of EPIC PN.
The images have been smoothed using a Gaussian filter.
The mosaic image shown in Figure \ref{mosaic} is a true color image
using the colors red for the 0.3 -- 0.8~keV band, green for 0.8 -- 1.5~keV, and 
blue for 1.5 -- 2.3~keV. 
The extended emission of the interstellar gas is clearly soft with no emission 
above $\sim$3~keV.
The comparison with the H$\alpha$ image in Figure \ref{hahi} shows
that the relatively bright extended region in the south of the PSR coincides
well with a superbubble in the \hii\ region N\,158 \citep{1956ApJS....2..315H},
which contains the OB association LH\,104 \citep{1970AJ.....75..171L}.
In order to study the spectral properties of the diffuse emission, we have 
selected two regions: region 1, which covers the brighter spot in the east of 
the PSR, and a region that covers the superbubble in the \hii\ region N\,158. 
The regions are shown in the left panel in Figure \ref{hahi}.
The PSR and the PWN with an extent of about 1\arcmin\ is completely 
removed from the data. The soft, extended emission east to the PSR
is not directly connected to the PWN and has, as we will see in Section
\ref{fits}, a perfectly thermal spectrum. Therefore, we assume that it is not 
related to the PWN.

\subsection{EPIC spectra}

\begin{figure*}[t]
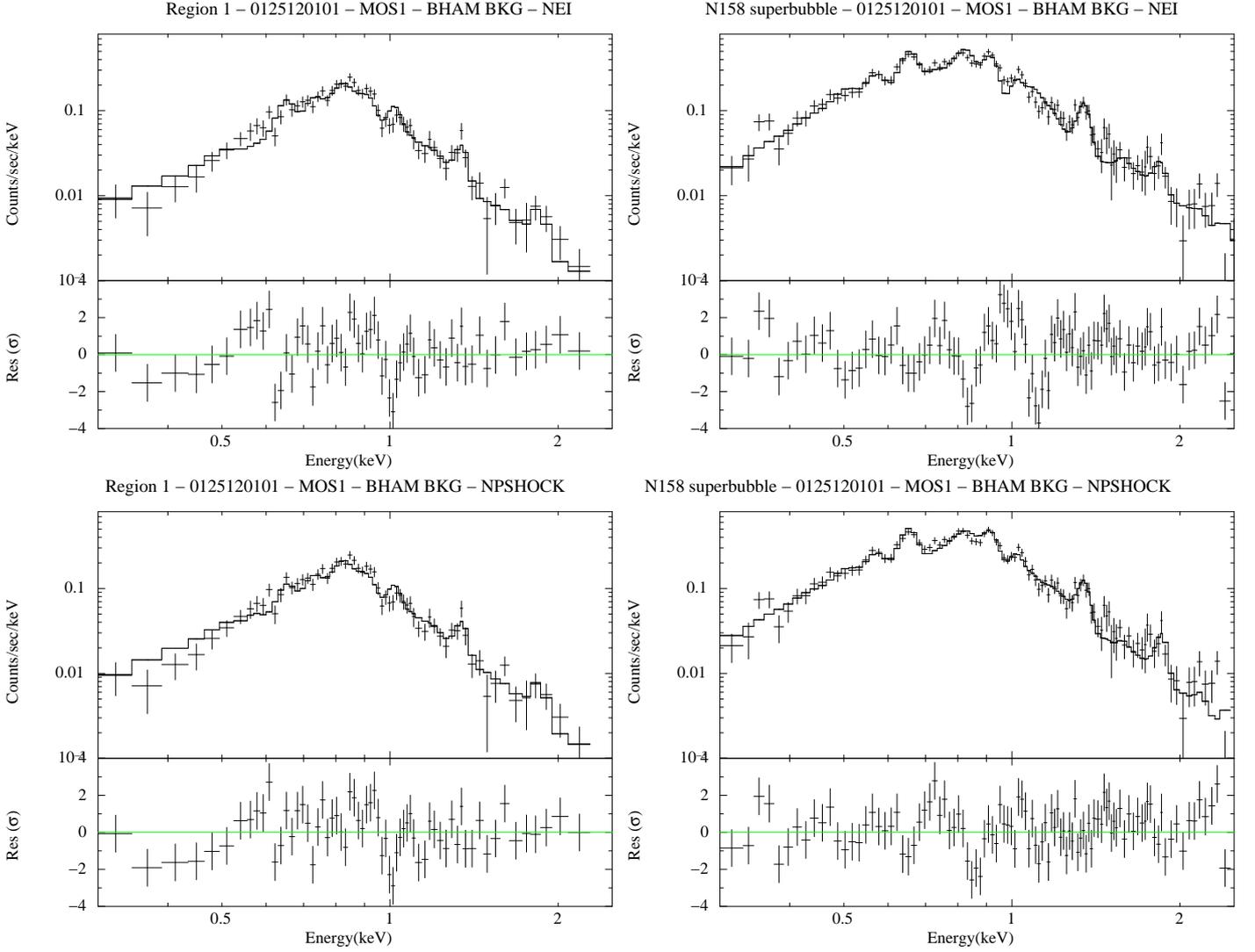

\includegraphics[width=0.39\textwidth,angle=270,clip=]{reg1_nei_MOS1.ps}
\hfill
\includegraphics[width=0.39\textwidth,angle=270,clip=]{n158_nei_MOS1.ps} \\
\includegraphics[width=0.39\textwidth,angle=270,clip=]{reg1_npshock_MOS1.ps}
\hfill
\includegraphics[width=0.39\textwidth,angle=270,clip=]{n158_npshock_MOS1.ps}\\
\caption{
Spectra of the diffuse emission in the field of view of PSR\,B\,0540$-$69
extracted from the \xmm\ EPIC MOS1 data of the Obs.\ ID 01251201. 
The left panels show the spectra of region 1 located east of the PSR,
the right panels show the spectra of the superbubble in the \hii\ region N\,158. 
}
\label{epicspec}
\end{figure*}

For the spectral anaylsis of an extended diffuse emission the contribution of 
the background is significant. As the emission
fills a large part of the detector, we are not able to extract a local 
background close to the source emission. We have to consider the following:
As the effective area of the mirrors depends on the off-axis angle, photons are 
subject to vignetting while particles are not.                     
The high-energy particles that interact with material surrounding the    
detector, however, produce fluorescence, which varies with position on the detector, 
especially for the PN detector.
In addition, the spectral response depends on the position on the detector.
A detailed description of the \xmm\ background is given by \citet{2003A&A...409..395R}
and \citet{2007A&A...464.1155C},
and a comparison of the different methods to estimate the background can be found
in, e.g., \citet{2004ApJ...617..322S}.

\subsubsection{Blank sky background}

The \xmm\ EPIC Background working group 
has created the so-called blank sky data for each EPIC and CCD read out mode
\citep{2007A&A...464.1155C}. The blank
sky data have been merged from data of different pointings after the point sources 
were eliminated from the data. These data sets 
comprise the detector background and an average cosmic X-ray background.   

Before extracting the spectra, we have first corrected both the observed data
and the blank sky data for vignetting using the XMMSAS command {\tt evigweight}.
The auxiliary response file (ARF) and the response matrix file (RMF) are then
created assuming that the source is on-axis. The background spectrum is extracted
from the blank sky data
at the same location on the detector as the source spectrum.

\subsubsection{Background from the same data}

For comparison, we also extracted a background spectrum from the same data as the 
source spectrum, using a region in the dark part north of B\,0540$-$69. 
However, after the subtraction of the background, the spectrum of the soft 
diffuse emission still has a hard tail and is over-corrected exactly at the energies 
of the fluorescence lines. Therefore, as expected, extracting the background from
the same data at a different position on the detector seems to be inappropriate.

\subsubsection{Closed filter wheel data}

Another way to deal with the background is to use a local background from the same
obervation, but take care of the detector background by using the closed filter wheel 
data, also supplied by the \xmm\ EPIC Background working group. For estimating the
X-ray background, i.e., additional emission that is typical for the observed area and 
might also contribute to the spectrum of the superbubble, we extract a region next the 
superbubble in the east, which shows faint diffuse emission. After subtracting the 
closed filter wheel spectrum extracted for each region at the corresponding position of 
the detector with the same shape, the spectrum of the faint diffuse emission in the east 
is fitted with a thermal model. This fitted spectrum is then included in the model of the 
spectra of the superbubble and region 1.

To verify if the emission in the east is suitable for the use as the local 
X-ray background, we estimate the flux for the superbubble, for the region east 
to it used as the local background, and 
for the blank sky background. We assume a plane-parallel shock model for the 
superbubble and for the eastern region a non-equilibrium ionisation (NEI) 
model, both with LMC abundances (see Sect.\,\ref{fits} for details). 
A combination of a thermal and a non-thermal spectrum is assumed for the blank 
sky background. From these spectra, we get the following fluxes with 90\% 
confidence errors in brackets: 
$F_{\rm 0.2-3.0~keV}$(superbubble) $= 1.4~(1.2 - 1.5) \times 10^{-12}$~erg~s$^{-1}$~cm$^{-2}$,
$F_{\rm 0.2-3.0~keV}$(east) $= 4.8~(4.0 - 5.3) \times 10^{-13}$~erg~s$^{-1}$~cm$^{-2}$,
$F_{\rm 0.2-3.0~keV}$(blank sky) $= 1.5~(1.4 - 1.6) \times 10^{-13}$~erg~s$^{-1}$~cm$^{-2}$.
The flux of the X-ray background in the blank sky data is about 10\% of that of the superbubble.
The emission from the ISM in the east is about three times higher than the blank sky. 
If we use the blank sky data for estimating the X-ray background, we might underestimate the
background. However, if we use the local emission, we will overestimate the background as 
we 
may expect hot gas also in the ISM next to the superbubble because of possible breakouts. 
The difference between the flux of the 
eastern region and the blank sky is about 20\% of the superbubble flux, which needs to be
taken into consideration as an additional uncertainty for the flux of the superbubble.

Otherwise, the results for the spectral fits obtained with the closed filter wheel data and 
those obtained with the blank sky data are consistent with each other within the confidence 
range of the spectral fit parameters. Therefore, in the following, we will discuss the 
fit results obtained with the blank sky data.

\subsubsection{Spectral fits}\label{fits}

The spectra are fitted with thermal plasma models in XSPEC. 
We fitted the spectra of the different EPICs simultaneously with model
parameters that are linked to each other.
In both spectra
emission peaks are found in the energy interval between 0.5~keV and 1.5~keV 
which can be interpreted as emission lines of highly ionised elements. 
The X-ray emission is absorbed by the Galaxy in foreground (\n{H\,fg}) and the
matter in the LMC in the line of sight (\n{H\,MC}) with abundances half of
solar values \citep{1992ApJ...384..508R}. 
The Galactic foreground column density is \n{H\,fg} $= 7.0 
\times 10^{20}$~cm$^{-2}$ in this region \citep{1990ARA&A..28..215D}.
The spectra are better reproduced by a NEI model 
\citep{2000RMxAC...9..288B} than by the spectral models assuming collisional
ionisation equilibrium (CIE). 
Figure \ref{epicspec} shows the MOS1 spectra of the Obs.\ ID 01251201
with the best fit models. In the following, 90\% errors are given for 
the parameters. 

\begin{table}                                
\centering                                                       
\caption{
Spectral parameters obtained from the fits of the EPIC data for region 1 and
the superbubble. The numbers in brackets are 90$\%$ confidence ranges. The parameter
$norm$ is the normalisation of the model in XSPEC and corresponds to the emission 
measure. For the other parameters, see Section \ref{fits}}
\begin{tabular}{lcc}                                                 
\hline\hline\noalign{\smallskip}
Parameter & \multicolumn{1}{c}{Region 1} & \multicolumn{1}{c}{Superbubble} \\
\noalign{\smallskip}\hline\noalign{\smallskip}
\multicolumn{3}{c}{NEI} \\
\noalign{\smallskip}\hline\noalign{\smallskip}
$N_{\rm H\,LMC}$ [$10^{22}$ cm$^{-2}$] & 0.14 (0.08 -- 0.23) &  0.43 (0.38 -- 0.51) \\
$kT$ [keV] & 0.54 (0.51 -- 0.58) & 0.91 (0.74 -- 1.1) \\
$\tau$ [$10^{11}$ s\,cm$^{-3}$] & 2.0 (1.6 -- 2.4) & 0.14 (0.12 -- 0.18) \\
$norm$ ($10^{-4}$) & 2.0 (1.6 -- 2.4) & 9.5 (8.0 -- 13.) \\
\noalign{\smallskip}
$\chi^2$/d.o.f & 94.81/61 = 1.55 & 185.06/106 = 1.75 \\
\noalign{\smallskip}\hline\noalign{\smallskip}
\multicolumn{3}{c}{NPSHOCK} \\
\noalign{\smallskip}\hline\noalign{\smallskip}
$N_{\rm H\,LMC}$ [$10^{22}$ cm$^{-2}$] & 0.19 (0.12 -- 0.32) &  0.29 (0.20 -- 0.33) \\
$kT_a$ [keV] & 0.55 (0.00 -- 0.59) & 1.0 (0.74 -- 1.2) \\
$kT_b$ [keV] & 0.51 (0.00 -- 0.59) & 0.13 (0.00 -- 0.57) \\
$\tau$ [$10^{11}$ s\,cm$^{-3}$] & 8.8 (5.4 -- 12.) & 1.3 (1.0 -- 2.3) \\
$norm$ ($10^{-4}$) & 5.0 (4.2 -- 6.3) & 11. (9. -- 13.) \\
\noalign{\smallskip}
$\chi^2$/d.o.f & 91.02/60 = 1.52 & 131.79/99 = 1.33\\
\noalign{\smallskip}\hline
\end{tabular}
\label{spectab}
\end{table}

Since the first fits with CIE yielded no satisfactory results, we use 
the NEI model in XSPEC with the effective temperature $kT$, the ionisation 
timescale $\tau = n_{\rm e}t $, and the abundance as fit parameters. 
The ionisation timescale $\tau$ is an indicator for the state of the
plasma after the gas has been ionised: For small ionisation timescales 
$\tau = n_{\rm e}t < 10^{12}$\,s\,cm$^{-3}$, the electrons and ions are 
still not in thermal equilibrium. Also for this model component, we use the
sub-solar LMC abundance $\zeta_{\rm LMC} = 0.5$. 
The fit of the spectra of the superbubble emission is still not very good 
with a reduced $\chi^2$ of 1.75. The emission is most likely caused by
the shock from the stellar winds in the superbubble. Therefore, we also fit 
the spectra using a plane-parallel shock plasma model with separate ion
and electron temperatures, $kT_a$ and $kT_b$, respectively 
\citep[NPSHOCK,][]{2001ApJ...548..820B}.
The fit parameters are given in Table \ref{spectab}.

As can be seen in Table \ref{spectab}, $kT_a$ and $kT_b$ for
region 1 are almost equal and the $\chi^2$ values for the NEI 
and the NPSHOCK fits are comparable. Therefore, the usage of
the NPSHOCK model does not seem to be necessary for region 1.
In addition, the value for $\tau$ is
$> 10^{11}$\,s\,cm$^{-3}$ for both NEI and NPSHOCK models, 
indicating that the gas in region 1 seems to be closer to CIE 
than in the superbubble in N\,158. 
The temperature of $kT =$ 0.54~keV determined 
for region 1 is higher and more accurate than the 
result of the \rosat\ PSPC analysis for this field 
\citep[0.3~keV,][]{2002A&A...392..103S}. 
This shows that we were only able to see the overall characteristic of a
larger region with PSPC, whereas with EPIC we can now resolve smaller 
(few arcminutes) structures of the hot ISM.

The hot plasma in the superbubble of the \hii\ region N\,158 seems 
not to be 
consistent with thermal equilibrium, indicated by the low ionisation 
timescale $\tau = 1.4 \times 10^{10}\rm{~s~cm}^{-3}$ of the NEI fit. 
The temperature of the NEI fit and the ion temperature of the NPSHOCK fit
are relatively high: $kT = 0.91\rm{~keV}$ and $kT_a = 1.0\rm{~keV}$,
respectively, while the electron temperature is one order of magnitude 
lower: $kT_b = 0.13\rm{~keV}$.  
Additionally to the Galactic foreground column density, a relatively high 
absorption column density of \n{H\,LMC,N158} $= 4.3$ or $2.9 \times 
10^{21}$~cm$^{-2}$ was determined. In comparision, the total column density 
in the LMC is \n{H\,LMC} $= 1.0 - 5.5 \times 10^{21}$~cm$^{-2}$ 
\citep{2005A&A...432...45B}. This corroborates that the diffuse X-ray 
emission most likely arises from inside the shell of the superbubble.

\section{Discussion}\label{discuss}

While in the east of the PSR\,B\,0540$-$69 the diffuse emission seems to arise 
from ionised gas close to thermal equilibrium with a temperature comparable to  
the value determined from the \rosat\ PSPC spectrum, in the south a smaller 
ionisation timescale and higher temperatures are determined. Since the emission 
coincides spatially with the superbubble in the
\hii\ region N\,158, we conclude that the origin of the diffuse X-ray emission 
is the hot gas within the interstellar bubble. The gas in the bubble 
interior is shocked by stellar winds, and the cooler outer rim is visible as 
an \hii\ region. The total unabsorbed X-ray luminosity of the bubble is 
$L_{\rm X}$(0.2 - 10.0~keV) = 1.5 $\times 10^{36}$~erg~s$^{-1}$.
Stars with masses over $25~M_{\odot}$ are luminous X-ray emitters with luminosities
of $L_{\rm X}$(0.2 - 10.0~keV) $\approx 10^{33}$~erg~s$^{-1}$.
In LH\,104 there are 16 stars with masses above 25~$M_{\odot}$ 
\citep[][see also Sect.\,\ref{anaest} for details]{1998A&AS..130..527T},
which account for 
$L_{\rm X}$(0.2 - 10.0~keV) $\approx 1 - 2 \times 10^{34}$~erg~s$^{-1}$.
This is two magnitudes lower than the emission from the bubble and can be neglected 
in the following discussion.

The comparison of the X-ray emission with the 
H$\alpha$ shell in Figure \ref{hahi} shows that 
there is additional X-ray emission outside the H$\alpha$ shell 
which might indicate that some hot gas escaped the superbubble. 
This has also been suggested by \citet{2001ApJS..136..119D} based on the analysis
of the \rosat\ data, which had already shown that the X-ray emission is not confined 
by the H$\alpha$ shell.
However, we are not able to rule out that some projection effect might make 
the H$\alpha$ look smaller than the extent of the X-ray emission.

\subsection{Results from the spectral analysis}\label{resspec}

For further studies, we use the NEI fit results for region 1 and NPSHOCK fit 
results for the superbubble.
The emitting volume can be approximated by an ellipsoid, although it is 
deformed in the south. We derive the radii from the EPIC mosaic image: $a =
4\arcmin\pm1\arcmin\ = (56\pm14)$~pc, $b = 3\arcmin\pm1\arcmin\ = (42\pm14)$~pc
($D_{\rm LMC} = 48$~kpc). We are not able to detemine the third radius of the
assumed ellipsoid. Therefore, we assume that the superbubble is oblate and
has a configuration like a disk perpendicular to the plane of the sky, i.e., 
also perpendicular to the disk of the LMC ($c = a$).
The volume of the bubble is then
$V = 4/3~\pi~abc = 
(1.62\pm0.03) \times 10^{61}~{\rm cm}^{3}$. 
With the LMC metallicity $\zeta_{\rm LMC} = 0.5$, $n_{\rm e} = (1.2 + 0.013 \zeta_{\rm LMC})~n
\approx 1.21~n$, 
with $n$ being the hydrogen density. 
Therefore, the normalisation of the spectral fit is
\begin{eqnarray}
norm & = & \frac{1}{10^{14} \times 4\pi D_{\rm LMC}^2} \int n_{\rm e} n {\rm d}~f~V \nonumber\\
& \approx& \frac{1}{10^{14} \times 4\pi D_{\rm LMC}^2}~1.21~n^{2}~f~V \nonumber\\
\nonumber\\
& = & 4.4 \times 10^{-62}~n^{2}~f~V \ \ [{\rm cm}^{-5}]
\end{eqnarray}
and the gas density within the bubble can be estimated as:
\begin{equation}\label{densinterior}
n = 4.8 \times 10^{30} \times \sqrt{\frac{norm}{f~V}} \ \ [{\rm cm}^{-3}].
\end{equation}
With $norm = (1.1\pm0.2) \times 10^{-3}$ from the fit with the NPSHOCK model, 
we get $n = (4.0\pm0.7) \times f^{-1/2} \times 10^{-2}~{\rm cm}^{-3}$.
If $f < 1$ then the density of $n = 4.0 \times 10^{-2}~{\rm cm}^{-3}$ is a 
lower limit. However, the angular resolution of the X-ray data does
not allow us to unambiguously determine the filling factor. For a 
young interstellar bubble like this particular case in N\,158 the filling factor 
can be assumed to be $f \approx 1$.

With temperature $T_a = (1.0\pm0.2)$~keV and density $n$ given, the pressure of
the gas is:
\begin{eqnarray}\label{pressure}
P/k & = & (n_{\rm e}+1.1 n) f^{-1/2} T_a = 2.31 n f^{-1/2} T_a \nonumber \\
& = & (1.1\pm0.3) \times f^{-1/2} \times 10^{6} ~{\rm cm}^{-3}~{\rm K}.
\end{eqnarray}
While the pressure of the Galactic ISM is thought to be
$P/k = 10^{3 - 4}~{\rm cm}^{-3}~{\rm K}$,
star forming regions in general have higher pressures of the order of
$P/k = 10^{5 - 6}~{\rm cm}^{-3}~{\rm K}$.
For a galaxy like the LMC with high star formation rate, 
\citet{2004ApJ...613..302O} estimated an ISM pressure of 
$P/k \approx 10^{5}~{\rm cm}^{-3}~{\rm K}$.
Thus the pressure inside the superbubble in N\,158 is about ten times higher 
than in the surrounding hot ISM. 
\citet{2005A&A...436..585D} have performed a 3-D simulation of 
the ISM including the effect of magnetic fields and obtained a map of the
distribution of temperature, pressure, magnetic field, etc. They have
shown that in regions where the temperature is about $10^{5 - 6}$~K, the 
pressure is $P/k = 10^{4 - 5}~{\rm cm}^{-3}~{\rm K}$, while it can reach
$P/k > 10^{5}~{\rm cm}^{-3}~{\rm K}$ in the interior of hot bubbles.
This is in agreement with the pressure that we obtain from the 
X-ray spectrum of the superbubble.

The part of the LMC in which N\,158 is located in general shows 
faint diffuse X-ray emission indicative of hot ISM. The region which we call 
region 1 is particularly bright and allows us to estimate the foreground
column density that causes absorption of the soft X-rays. 
The shell of cooler gas around the superbubble should make an additional 
absorbing component. 
Using the absorbing column density determined for the region 1 
(besides the Galactic column density) as the mean LMC value 
\n{H\,LMC,region1} $= (1.4\pm0.9) \times 10^{21}$~cm$^{-2}$, the column 
density of the shell around the superbubble is
$N_{\rm H\,shell} = N_{\rm H\,LMC,N158} - N_{\rm H\,LMC,region1} = 
(1.5\pm1.3) \times 10^{21}~{\rm cm}^{-2}$.

\subsection{Analytic estimates}\label{anaest}

With the values derived from spectral fitting for pressure and density inside 
the bubble, we infer a thermal energy content of 
$E(t) = 3/2 \times P \times f \times V = 3.6 \times f^{1/2} \times 10^{51}$~erg.
We estimate a mean shell thickness of $\sim$0\farcm5, i.e., $\sim10\%$ of $a$, 
corresponding to $\sim$7~pc from the MCELS H$\alpha$ image shown in 
Figure\,\ref{hahi}.
From the column density $N_{\rm H\,shell}$ calculated in the last section,
we obtain a density of 
$n_{\rm{shell}} \approx 70\,$cm$^{-3}$ 
inside the shell.
Since the ISM around N\,158 shows H$\alpha$ emission, its temperature is 
probably around 8000~K, typical for the warm ionised medium 
\citep{1977ApJ...218..148M}, resulting in a speed of sound of $\sim$~9.2~km/s.
With the inferred pressure 
of $P/k = 1.1 \times f^{-1/2} \times 10^{6} ~{\rm cm}^{-3}~{\rm K}$
from equation \ref{pressure}
we obtain a density of the ambient medium of 
$\sim$13~cm$^{-3}$, yielding a compression factor of the shock of $\sim$6. 
Since the cooling time behind the shock is very small due to the high density in
the shell, the shock will be isothermal and, as shown below, also strong.
Since we now know the radius, energy content, and density, we can use 
the solution for a wind blown bubble by \citet{1977ApJ...218..377W} to
find out the age of the bubble, which is about 
$t \simeq 1.1$~Myr
resulting in an energy input rate over this time interval of 
$L_{\rm{superbubble}} = 2.3 \times 10^{38}$~erg~$s^{-1}$.
In addition, the mass inside a homogeneous bubble is
$M = 2.31 n \times \mu \times m_H \times V = 770$ M$_{\odot}$,
where $\mu=0.61$ is the mean molecular weight of a fully ionised gas 
and $m_H$ is the hydrogen mass.
Thus, a mass loss rate of 
$6.9 \times 10^{-4}$ M$_{\odot}$/yr
over 
1.1~Myr
is derived.
In the following we will discuss if massive stars can account for such a large 
mass loss and energy input rate.

In order to calculate the mass loss and energy input rates of OB stars (in cgs units) 
we use the mass-luminosity relation for stars with $10 \le M/M_{\odot} \le 50$
by \citet{2007AstL...33..251V} to obtain the mass $M$ and luminosity $L_{\star}$ 
of a star with bolometric magnitude $M_{\rm{bol}}$:
\begin{equation}
M_{\rm{bol}}=1.6-6.9 \log (M/M_{\odot})
\end{equation}
\begin{equation}
L _{\star}= 19 (M/M_{\odot})^{2.76} L_{\odot}
\end{equation}
The radius of the star is obtained from
\begin{equation}
R = \sqrt{\frac{L_{\star}}{4\pi\sigma  T_{\rm{eff}}^4}}
\end{equation}
where $T_{\rm{eff}}$ is the effective temperature of the star and $\sigma$ is 
the Stefan-Boltzmann constant.
$M_{\rm{bol}}$ and $T_{\rm{eff}}$ are taken from Table 2 of \citet{1998A&AS..130..527T}.
The wind velocity is determined according to the theory of radiation-driven winds 
\citep{1975ApJ...200L.107C}:
\begin{equation}
v_{\infty}= a v_{\rm{esc}}=a \left[ \frac{2GM}{R} \times (1-L_{\star}/L_{\rm{edd}})\right]^{0.5}
\end{equation}
where $a \simeq 2.5$ \citep{1995ApJ...455..269L}. The parameter $v_{\rm{esc}}$ is the 
photospheric escape velocity and the Eddington luminosity is
$L_{\rm{edd}}=4 \pi G \times M \times m_p \times c/\sigma_T$ with $m_p$ being the
mass of a proton and $\sigma_T$ the Thomson cross-section for the electron.
The mass loss is determined from the single-scattering limit
\begin{equation}
\dot{M}=\frac{L_{\star}} {v_{\infty} \times c} \, .
\end{equation}
We corrected for the LMC metallicity $\zeta_{\rm{LMC}}$ following 
\citet{1992ApJ...401..596L} and obtain $\dot{M}\propto \zeta_{\rm{LMC}}^{0.8}$ for 
the mass loss and $v_{\infty} \propto \zeta_{\rm{LMC}}^{0.13}$ for the wind velocity 
(for hot stars with $M > 15 M_{\odot}$).
Thus, the 67 O- and B-stars in LH\,104 produce a mass loss of 
$\sim$ 39 $M_{\odot}$
or an energy input rate of 
$L_{\rm{OB}}=3.4 \times 10^{37}$~erg~$s^{-1}$
over 
1.1~Myr.

Additionally, the WR-stars generate $\sim$ 35 $M_{\odot}$ or 
$L_{\rm{WR}}=1.8 \times 10^{38}$~erg~$s^{-1}$, but only for
$2.5 \times 10^5$~yr~$\cong0.23~t$, assuming that they already 
went through half of their WR-lifetime of $\sim5\times 10^5$~yr 
\citep{1987A&A...182..243M}.
Mass loss and wind velocity of WR-binaries are adopted from \citet{1997ApJ...481..898L}
and corrected for mass losses of WC and WN types in the LMC according to 
\citet{2007ARA&A..45..177C}.
Furthermore, we have to calculate the contribution of winds from WR-binaries 
before entering the WR-phase, i.e., for the remaining 0.85~Myr.
With mass losses for O6 and O7-stars and O4-stars as WR-progenitors (PR) taken from 
\citet{1981ApJ...250..660G} and velocities from \citet{1993ApJ...412..771L}, we get
4~$M_{\odot}$ or $L_{\rm{PR}} = 8.0 \times 10^{36}$~erg~$s^{-1}$ for
0.85~Myr~$\cong0.77~t$
from these stars.

In total, we find that winds can account for 
$\sim78~M_{\odot}$ or a mechanical luminosity of
$L = 8.0 \times 10^{37}$~erg~$s^{-1}$ over 1.1~Myr.
Still, the largest part of the mass 
($\sim$690~$M_{\odot}$) or 
$L_{\rm{superbubble}} - L = 1.5 \times 10^{38}$~erg~$s^{-1}$
corresponding to a thermal energy of $2.3 \times 10^{51}$~erg
is `missing', but this can be partly explained 
after applying an IMF to the star cluster. 
According to the Hertzsprung Russell diagram (HRD) of 
Testor \& Niemela (\citeyear{1998A&AS..130..527T}, Fig. 6b) and using the masses 
derived from the $M$-$L$ relation \citep{2007AstL...33..251V}, we find that 16 stars 
in LH\,104 have masses above 25~$M_{\odot}$ including all O-stars,
WR-binaries, the B0V stars with Id 4-41 and 4-55, and Sk-69 259. As an upper mass limit 
we take $65~M_{\odot}$ as a rough estimate, since \citet{2000AJ....119.2214M} 
suggested that the progenitor masses of WR-stars in this cluster should be in excess of 
$60~M_{\odot}$. On the other hand, by looking at the HRD of LH\,104 in 
Massey et al.\,(\citeyear{2000AJ....119.2214M}, Fig.7) gives at least 20 stars in the 
25 -- 65 $M_{\odot}$ interval, for which spectral types or photometry are available.
Comparing both HR-diagrams shows that the total number of member stars is quite 
uncertain, but there must be at least 70 stars above 8~$M_{\odot}$.

Assuming that all stars formed from the same parental cloud and using an IMF with 
$\Gamma=-1.05$ \citep{1998A&AS..130..527T} we obtain a total of 74 stars between
8 and 65~$M_{\odot}$ in the first case (16 stars with $25 < M/M_{\odot} < 
65
$)
and 93 stars in the other case (20 stars with $25 < M/M_{\odot} < 
65
$). In both cases there are about 2 stars in the 65--80 $M_{\odot}$ mass interval
or about 3 stars in the 65 -- 90 $M_{\odot}$ mass interval.
This suggests that 2 -- 3 supernovae (SNe) already exploded around 1~Myr ago and 
with $E_{\rm{SN}}=10^{51}$~erg per SN explosion, they can easily account for the 
required amount of energy.
These 
2 -- 3 SNe
yield approximately 
150 -- 240~$M_{\odot}$
of ejecta mass, thus 
230 -- 310 of 770
solar masses in the bubble can be explained. 
Remaining discrepancies should be due to mass loading and evaporation of 
entrained interstellar clouds and/or turbulent mixing of material from the cold 
shell.

With the parameters determined so far we can further investigate the geometry of the 
bubble and the ambient ISM.
We use an analytical model for the expansion of a wind-blown bubble in an exponentially 
stratified medium symmetric to the galactic midplane (Baumgartner $\&$ Breitschwerdt, 
in prep.) based on the approximation of \citet{1960SPD.....5...46K}.
With an energy input rate of 
$L_{\rm{superbubble}}=2.3 \times 10^{38}$~erg~$s^{-1}$
the bubble should reach a radius, i.e., semi-major axis, of 56~pc after 
$\sim$1.1~Myr.
We calculated 
models with different scale heights ($H=50, 100$, and $500$~pc) and obtain a 
density of the ambient medium of 
$n_0= 9, 11$, and $13$~cm$^{-3}$,
respectively.
Values for the height of the bubble above/below the galactic plane are $c=80, 66$, and 
$58$~pc. Using $b=42$~pc, we obtain a volume of the bubble of 
$V=2.4 \times 10^{51}$~cm$^3$, $1.9 \times 10^{61}$~cm$^3$ and 
$1.7 \times 10^{61}$~cm$ ^3$. Since a larger elongation, i.e., lower scale height, 
yields a larger volume, we argue that a larger scale height fits better, otherwise 
resulting in a density inside the bubble that is too low.
With $V=1.7 \times 10^{61}$~cm$^3$ and $norm$ from Sect.\,\ref{resspec}, 
we derive 
$n=3.9 \times 10^{-2}$~cm$^{-3}$.
This is close to the value that was used to 
calculate the mass inside the bubble and the energy input rate in the beginning, 
whereas a volume of $V=2.4 \times 10^{51}$~cm$^3$ for $H=50$~pc yields only 
$n=3.2 \times 10^{-2}$~cm$^{-3}$.
The shell thickness in the case of $H=500$~pc is $\sim5$~pc resulting in a density of 
the shell of 100~cm$^{-3}$, which is 
somewhat higher than the density of 70~cm$^{-3}$ derived from the observation.
Finally, we calculated the velocity of the outer shock,
which propagates into the ambient warm medium
at the top/bottom of the bubble
of 31~km~s$^{-1}$,
which agrees
very well
with the expansion velocity found with
the wind solution of \citet{1977ApJ...218..377W}
and gives a Mach number of $M \sim 3.4$.
For comparison, 
\citet{2001ApJS..136..119D} 
reported 
an expansion velocity 
of the superbubble of $v_{\rm exp} \approx 45$~km~s$^{-1}$ assuming a 
pressure-driven bubble in an homogeneous medium. 
The magnetic fields in the ISM of
the LMC are on the order of $\sim1 \mu$G \citep{2005Sci...307.1610G}
corresponding to an Alfv\'en velocity of $\sim$0.5~km~s$^{-1}$. Therefore,
the effects of the magnetic fields are negligible.

\section{Summary}

We have found significant diffuse X-ray emission in the field of view of the 
\xmm\ observations of the pulsar B\,0540--69, which is not related to the pulsar.
The analysis of the spectrum of the emission has shown that it is
purely thermal and can be best modelled with a hot shocked gas. The extended diffuse
emission coincides spatially with the superbubble in the \hii\ region N\,158 in the 
LMC.
Therefore, we conclude
that the origin of the X-ray emission is the hot gas inside the superbubble in N\,158.

From the parameters obtained from the analysis of the \xmm\ spectra we derive a
temperature of $kT = 1$~keV, a density of $n = 0.04$~cm$^{-3}$, and a pressure of 
$P/k = 10^{6}$~cm$^{-3}$~K inside the superbubble. These results enable us
to perform analytic calculations of the evolution of the superbubble.
Since the OB association LH\,104 that is located inside the superbubble has been 
studied in detail \citep{1998A&AS..130..527T}, the stellar population inside the 
superbubble is well known. Using its IMF and HRD we calculate the mass loss and 
energy input rates of the stars. We estimate an age of
$\sim$~1~Myr
and a total energy input rate of
$L = 2.3 \times 10^{38}$~erg~s$^{-1}$.
The massive stars including WR-stars and binaries in LH\,104 account for 
$L = 8 \times 10^{37}$~erg~s$^{-1}$.
Therefore, in order to reproduce the observations, there should have been
2 -- 3
SN explosions in the past
1~Myr.
We have also
performed calculations of the expansion of the superbubble for different scale heights.
We show that a large scale height of $H = 500$~pc can well reproduce the observed 
density inside the superbubble as well as the density in the shell around it. 
The corresponding expansion velocity of the superbubble is also in 
good agreement with the model of \citet{1977ApJ...218..377W}.

\begin{acknowledgements}
The authors thank the anonymous referee for comments that helped to improve the paper.
This research is based on observations obtained with \xmm, an ESA science mission 
with instruments and contributions directly funded by ESA Member States and NASA.
We have also made use of preliminary data of the UM/CTIO Magellanic Cloud Emission Line
Survey (MCELS) availabel on the web site of the project 
({\it http://www.ctio.noao.edu/mcels/}). 
M.S.\ acknowledges support by the Deutsche Forschungsgemeinschaft through the Emmy 
Noether Research Grant SA 2131/1.
V.B.\ is recipient of a DOC-fFORTE fellowship of the Austrian Academy
of Sciences.
\end{acknowledgements}

\bibliographystyle{aa} 
\bibliography{../../bibtex/xraytel,../../bibtex/my,../../bibtex/nearbygal,../../bibtex/ctb109,../../bibtex/ism.bib}


\end{document}